# PROXY DISCRIMINATION AFTER *STUDENTS FOR FAIR ADMISSIONS*

FRANK FAGAN*

Law protects people from discrimination. Algorithms, however, can easily circumvent the appearance of discrimination through the artful use of proxy variables. For instance, a lending algorithm may appear to satisfy a legal standard by ignoring race, but the same algorithm might deny loan applicants on the basis of having attended a particular high school—a variable that may closely correlate with race. An algorithm that assesses work performance and recommends promotions may ignore sex, but the same algorithm might penalize employees who take, on average, more paternity leave—a variable that may closely correlate with sex. The abuse of proxies cuts across political views on affirmative action. For example, an admissions committee might technically ignore race consistent with recent changes to Equal Protection rules, but the same committee might consider variables that are highly correlated to race, such as zip code, high school, and the income of parents, in order to achieve a university's diversity goals. Today, there is no clear legal test for regulating the use of variables that proxy for race and other protected classes and classifications. This Article develops such a test. Decision tools that use proxies are narrowly tailored when they exhibit the weakest total proxy power. The test is necessarily comparative. Thus, if two algorithms predict loan repayment or university academic performance with identical accuracy rates, but one uses zip code and the other does not, then the second algorithm can be said to have deployed a more equitable means for achieving the same result as the first algorithm. Scenarios in which two algorithms produce comparable and non-identical results present a greater challenge. This Article suggests that lawmakers can develop caps to permissible proxy power over time, as courts and algorithm builders learn more about the power of variables. Finally, the Article considers who should bear the burden of producing less discriminatory alternatives and suggests plaintiffs remain in the best position to keep defendants honest—so long as testing data is made available.

* Associate Professor, South Texas College of Law Houston; Research Associate, EDHEC Augmented Law Institute (France). Comments welcome: ffagan@stcl.edu. For comments, I thank fellow panelists and participants of the Affirmative Action and Algorithmic Fairness symposium held at Stanford Law School on May 19, 2023, participants of the law and economics workshop held at the Coase-Sandor Institute, University of Chicago Law School on March 19, 2024, as well as Saul Levmore, Ryan H. Nelson, and Julian Nyarko.

## INTRODUCTION

In its recent landmark ruling on affirmative action, *Students for Fair Admissions (SFFA) v. Harvard*, the U.S. Supreme Court held that race-based university admissions programs violate law if they "lack sufficiently focused and measurable objectives," "unavoidably employ race in a negative manner, involv[ing] racial stereotyping," and "lack meaningful end points."[1] The holding is expected to dismantle affirmative action in college admissions.[2] But curiously, within the same opinion, the Court noted that a university may permissibly consider how race has affected an applicant's life, "be it through discrimination, inspiration, or otherwise."[3] The idea is that race may lead to unique experiences that shape a person as an individual and instill a desirable attribute or set of attributes in the process. For instance, a Native American growing up in contemporary America may have faced and overcome discrimination. The experience of surmounting discrimination may have granted that person courage, a desirable attribute for a potential student. It is possible to consider the effects of race here without considering race itself and engaging in racial stereotyping because focus can be placed on courage.[4]

[1] Students for Fair Admissions, Inc. v. President & Fellows of Harv. Coll. (*SFFA*), 600 U.S. 181, 230 (2023).

[2] *See, e.g.*, Jelani Cobb, *The End of Affirmative Action*, NEW YORKER (June 29, 2023), https://www.newyorker.com/magazine/2023/07/10/the-end-of-affirmative-action.

[3] *SFFA*, 600 U.S. at 230.

Of course, the university could be dishonest. It could ignore the applicant's connection of race to courage and choose instead to focus only on race. When called upon to give account of its decision, the university could attempt to reconstruct its decision-making process as one that was focused on courage. A factfinder would be asked to evaluate the evidence and make a decision, as in other cases with difficult evidentiary records. Today, as might be expected, such dishonesty is unlawful. The *SFFA* Court noted as much:

> [W]hat cannot be done directly cannot be done indirectly. The Constitution deals in substance, not shadows, and the prohibition against racial discrimination is leveled at the thing, not the name.[5]

As highlighted by the *SFFA* dissent, colleges and universities can continue to recruit on the basis of race-neutral factors such as socioeconomic status, the ability to speak multiple languages, and whether one's parents went to college.[6] These factors are, on their face, race-neutral despite the fact that they correlate to race. Across the United States, lower socioeconomic status is correlated with being African American. The ability to speak multiple languages correlates with being Hispanic. Of course even a casual observer of statistical science knows that correlation is not causation. Thus, statisticians look for mediating variables, that is, variables that transmit causation. For example, being African American causes a person to experience discrimination on average, which causes the socioeconomic status of African Americans to be low on average in turn. In this way, being African American is associated with, or co-related to, low socioeconomic status. Both before and after *SFFA*, universities can lawfully recruit on the basis of race-neutral variables even if they are correlated with race.[7]

---

[4] If the university admits the applicant solely on the basis of Native American ancestry because the school believes all Native Americans are courageous, then it would be employing a racial stereotype. It does not matter that the stereotype is "positive", that is, it casts all Native Americans in a positive light by inferring that they all possess courage. By using race in a positive manner, applicants of other races are excluded in higher numbers. Admitting more Native Americans on the basis of a stereotype leads to fewer applicants being admitted from other racial groups. Instead, if all applicants are evaluated on the basis of courage, then either more or less of each group could be granted admission on the basis of a neutral attribute that all applicants potentially possess.

[5] *SFFA*, 600 U.S. at 230-31 (quoting Cummings v. Missouri, 71 U.S. 277, 325 (1867)) (internal quotation marks omitted).

[6] *Id.* at 231.

[7] *Cf.* Village of Arlington Heights v. Metro Housing Development Corp., 429 U.S. 252 (1977) (finding zoning ordinance that barred construction of multi-family dwellings

Now a clever admissions committee that understands how a variable correlates with race can achieve its diversity objectives indirectly. Note that this method cuts both ways. Indirect variables can be used to promote or thwart educational diversity. Suppose a particular variable combination, say, high standardized test scores and hours of volunteer work, are correlated with being East Asian. A college can intentionally suppress East Asian admissions, without considering race, by disfavoring this particular combination variable.[8] Such a strategy is unlawful. To paraphrase the Court, what cannot be done directly cannot be done indirectly by means of race-neutral variables that correlate with race. But this rule presents a puzzle of sorts. As hinted at above, it is evidentiary in character. The clever admissions committee may lawfully consider the combination variable as long as its consideration is not intended to deny admission to East Asians. It additionally seems to present a legal issue involving the degree of correlation power. If a variable correlates strongly and conspicuously to race, but does little to achieve a lawful goal, how can a decision-maker credibly use it in innocence?[9]

Consider a bank whose internal lending model predicts higher default rates for divorcees. The law prohibits the denial of loans on the basis of marital status,[10] but suppose the bank observes two associated variables: change in last name and number of joint bank accounts.[11] The bank may

---

permissible because no direct discriminatory purpose was found, even though the ordinance indirectly discriminated by reducing the availability of affordable and accessible housing to people of different socio-economic and ethno-racial backgrounds).

[8] Statisticians will recognize the combination of two variables as yielding "interaction variables." *See* JEFFREY M. WOOLRIDGE, INTRODUCTORY ECONOMETRICS: A MODERN APPROACH 240-41 (5th ed. 2013). Two variables are multiplied together to produce a third variable, which attempts to capture the impact of the interaction on a dependent variable of interest. For example, suppose I am interested in predicting academic success by using the two variables mentioned above: standardized test scores and hours of volunteer work. After computation, I find that my two-variable model explains 20 percent of the variation in academic success. Not satisfied, I construct a third variable by multiplying together standardized test scores and hours of volunteer work. After computation, I find that my three-variable model explains 50 percent of the variation in academic success. When adding the third "interaction" variable, the model explains a greater amount of the variability in academic success. It is a stronger model.

[9] The logic follows that of the yarmulke-tax analogy made in Bray v. Alexandria Women's Health Clinic, 506 U. S. 263, 270 (1993):

> Some activities may be such an irrational object of disfavor that, if they are targeted, and if they also happen to be engaged in exclusively or predominantly by a particular class of people, an intent to disfavor that class can readily be presumed. A tax on wearing yarmulkes is a tax on Jews.

[10] *See* 15 U.S.C. § 1691(a)(1) (declaring marital status as a protected classification).

[11] Proxies are everywhere. For example, many variables correlated to race are used in lending decisions. *See* Neil Bhutta, Aurel Hizmo, and Daniel Ringo, *How Much Does Racial*

lawfully consider these seemingly innocuous variables as long as their consideration is not intended to deny loans to divorcees or otherwise produce an unlawful disparate impact. But these variables do not predict default on their own; they do so through their association with marital status.[12] Suppose the lending model explains the variability in the likelihood of loan default with the two variables only. And suppose in this example that the model explains 100 percent of the variability in default probability; fifty percent is attributed to the change in last name, and fifty percent is explained by the number of joint bank accounts. Now, a careful statistician adds a third variable, marital status. After carrying out diagnostic testing of the new model, the statistician learns that neither change in last name nor number of joint bank accounts explains any amount of variability in loan default probability. Marital status explains it completely, and the predictive powers of the innocuous variables fall to zero. In this toy example, the two original variables are *proxies* for marital status.[13] Once the protected-classification variable is added to the model, the predictive power of the proxies disappears. Should the bank be permitted to make lending decisions on the basis of these variables?

One might be tempted to say no. After all, the proxy variables in the example are statistically identical to the protected-classification variable.[14] However, by the terms of relevant law, the bank may permissibly treat people of a protected class differently from others when accounting for credit risk factors.[15] Thus, the question, as strange as it may seem, is whether change in last name and number of joint bank accounts represent credit risk factors. Directly, they do not. Yet indirectly, they are associated with divorce, which is associated with elevated credit risk in turn.

Fair lending law thus deploys a broader prohibition on disparate impacts. As is well known, a facially neutral "policy or practice" that has an adverse impact, such as assessing credit risk on the basis of change in last

*Bias Affect Mortgage Lending? Evidence from Human and Algorithmic Credit Decisions*, Fed. Res. Board Working Paper 2022-067 (2022).

[12] Divorced people tend, on average, to default on their loans more than married people. *See* Kristopher Gerardi, et al., *Can't Pay or Won't Pay: Unemployment, Negative Equity, and Strategic Default*, NBER Working Paper No. 21630 (Oct. 2015) at 61 (providing statistical evidence of the phenomenon).

[13] For further examples, see James Grimmelmann & Daniel Westreich, *Incomprehensible Discrimination*, 7 CALIF. L. REV. ONLINE 164 (2017); Darcy Steeg Morris, Daniel Schwarcz & Joshua C. Teitelbaum, *Do Credit-Based Insurance Scores Proxy for Income in Predicting Auto Claim Risk?*, 14 J. EMPIRICAL LEGAL STUD. 397, 418–21 (2017).

[14] More precisely, they explain the identical level of variation in default rate.

[15] Explanatory notes to the applicable regulation provide the example of marital status explicitly: "The fact that certain credit-related information may indirectly disclose marital status does not bar a creditor from seeking such information." 12 C.F.R. Pt. 1002, Supp. I, Sec. 1002.5(d)(1).

name or number of joint bank accounts, cannot be used unless it serves a legitimate business need that "cannot be achieved by means that are less disparate in their impact."[16] To set the stage for the core issue, two assumptions can be made. First, assume that the selection of variables for a lending model is a "policy or practice."[17] Second, assume that minimizing credit risk is a "legitimate business need" of a bank.[18] Thus, the lawful use of the two offending variables turns on the availability of other variables that are unrelated (or less strongly related) to marital status, but nonetheless hold the same predictive power. For instance, suppose the statistician adds a fourth variable to the model—the loan applicant's history of military service. After diagnostic testing, the statistician learns that across the population of loan applicants, military service history is entirely unrelated to marital status. He removes the two facially neutral but offending variables and adds the military service variable. If this basic model continues to explain 100 percent of the variability in loan default probability, then the statistician has clearly found an alternative method for achieving the bank's legitimate business need.[19]

---

[16] 12 C.F.R. Pt. 1002, Supp. I Sec. 1002.6(a)-2.

[17] This assumption was recently made in the context of employment selection procedures by the EEOC. *See* U.S. Equal Emp. Opportunity Comm'n, *Select Issues: Assessing Adverse Impact in Software, Algorithms, and Artificial Intelligence Used in Employment Selection Procedures Under Title VII of the Civil Rights Act of 1964* (May 18, 2023), https://www.eeoc.gov/laws/guidance/select-issues-assessing-adverse-impact-software-algorithms-and-artificial#_edn6.

[18] Repayment of loans is generally recognized as one. 12 C.F.R. Pt. 1002, Supp. I Sec. 1002.6(a)-2, which notes:

> For example, requiring that applicants have income in excess of a certain amount to qualify for an overdraft line of credit could mean that women and minority applicants will be rejected at a higher rate than men and nonminority applicants. If there is a demonstrable relationship between the income requirement and creditworthiness for the level of credit involved, however, use of the income standard would likely be permissible.

[19] This discovery is probably unlikely inasmuch as marital status is highly predictive of default rate in this example. If military service explains default rate, then the variable is probably correlated with marital status. The general point is that some models are better than others even if they are hard to imagine. Legal scholars try to imagine better models that are also consistent with existing law, where "better" often means more equitable in the short- and long-term. *See* Aziz Z. Huq, *Racial Equity in Algorithmic Criminal Justice*, 68 DUKE L.J. 1043, 1094-95 (2019) (suggesting that a model's desirability should be measured on the basis of its long-term effects on racial stratification); Pauline T. Kim, *Race-Aware Algorithms: Fairness, Nondiscrimination, and Affirmative Action*, 110 CALIF. L. REV. 1539, 1544-45 (noting that models should be de-biased, and that some methods for de-biasing algorithms do not constitute discrimination and therefore deserve no scrutiny). *Compare* Jason R. Bent, *Is Algorithmic Affirmative Action Legal?*, 108 GEO. L.J. 803, 825-41, 852 (2020) (suggesting that race aware de-biasing strategies require standard affirmative action justifications); Daniel E. Ho & Alice Xiang, *Affirmative Algorithms: The Legal Grounds for Fairness as Awareness*, 2020 U. CHI. L. REV. ONLINE 134, 136 (2020) (noting that de-biasing

And if the new model suggests making a loan to just one more divorcee than the old model, then it is less disparate in its impact.

A general rule can be drawn from this example. Proxies for a protected class or classification should be removed if their removal leads to no loss to a model's accuracy. This is the basic insight of this Article. What if the replacement of proxy with non-proxy variables reduces accuracy in predicting loan defaults? If the reduction in discrimination is large and accuracy losses are small, then it would appear to be socially desirable to prohibit the use of the proxies in this instance. The Article is mostly agnostic with respect to this question and takes whether a goal is lawful as given.[20] In other words, once law determines that a goal is socially desirable, universities, employers, and lenders are free to achieve it so long as their efforts are narrowly tailored or otherwise represent a less discriminatory alternative. Nonetheless, the idea of proxy power caps, elaborated in Part III, provides some room for reducing proxy discrimination when accuracy losses appear to be small. Comparatively small losses very likely represent differences in measurement technique and are more efficiently resolved with a straightforward cap. Thus, the tradeoff is resolved on the basis of administrative cost reduction.

Functionally, disparate impact and narrow tailoring rules accomplish the same thing. They encourage a discriminating party to use better means to achieve a goal if better means are available. Of course the goal must be socially reasonable, that is, it must represent a "legitimate business need" or a legitimate "interest" of government. Society permits its judges to police whether the goals of a discriminating party are socially reasonable. Discerning the reasonableness of a goal may be difficult, especially when it is politically charged, but its discernment requires no great technical effort. It is a moral decision.

By contrast, whether better means are available is more difficult to assess technically, especially in the context of algorithms and other complex methods for goal achievement. While no great ethical acumen is required, a judge must assess the means among a universe of alternatives. Suppose the

---

strategies necessarily consider race and other protected classes even if potentially inconsistent with affirmative action law); Sandra G. Mayson, *Bias in, Bias Out*, 128 YALE L.J. 2218, 2230, 2262–63 (2019) (noting that using race in a model may constitute intentional discrimination).

[20] The legality of some goals is more settled than others. For instance, credit risk minimization is lawful, and the use of race-neutral variables is permitted within limits. In contrast, the lawful pursuit of educational diversity or other race-related goals with facially neutral selection criteria has recently been questioned after *SFFA*. *See* Coal. for TJ v. Fairfax Cnty. Sch. Bd., 68 F.4th 864, 871 (4th Cir. 2023); *see also* Sonja Starr, *The Magnet School Wars and the Future of Colorblindness*, 76 STAN. L. REV. 161, 161 (2024) (providing argument in favor of lawfulness).

bank in the above example is unaware that history of military service can effectively suppress the need for using marital status proxies. In our adversarial system, we generally place the burden on the divorcee to discover the better means.[21] It is unclear whether this will continue to make sense as banks and other defendants may be in the best position to evaluate their algorithms (and other complex and private means of decision-making),[22] though it is hard to imagine how a defendant could be incentivized, let alone trusted, to carry out a costly search for alternatives. In the race-based admissions context, the defendant must always narrowly tailor her means. Incentives are therefore less of a problem. But even here, when should the defendant permissibly be able to stop searching for better alternatives? This point will be taken up in Part IV.

Part I provides an overview of anti-discrimination law in the admissions, lending, and employment contexts. Emphasis is given to proxy issues throughout the discussion. Part II lays out a precise definition of a proxy and develops the concept of an algorithm's total proxy power. As mentioned above, a less discriminatory alternative exhibits comparatively less proxy power than its counterparts. Part II explains the intuition for how proxy power can be measured. Part III discusses how the law can deploy the proxy power concept to reduce indirect discrimination. Several approaches are examined, including outright prohibition of proxy variables, a comparative minimum proxy power standard, and a set cap on the total proxy power. Finally, Part IV turns to who should bear the burden of searching for alternatives, and suggests that plaintiffs remain in the best position to discover indirect discrimination, so long as they can access sufficient data for testing.

## I. ANTI-DISCRIMINATION LAW

Resources are not infinite. Usually, they are purposefully allocated to effectively achieve goals. In some instances, resources can be distributed randomly, as is done in citizenship lotteries, and goals can still be achieved.[23] Although random distribution can eliminate intentional discrimination and provide other benefits,[24] it often reduces the ability of a resource holder to satisfy an objective. If, for example, an open position for a fashion designer were allocated randomly, then a fashion house would hire a sufficiently skilled designer only if it drew one from the population of applicants by chance. To move toward its goal of good fashion design, the fashion house would need to be lucky. On the other hand, any time a holder of scarce resources makes a non-random allocation, there is a chance of disparate impact.[25] If the fashion house were to hire the top ten designers in terms of ability as ranked by its hiring algorithm, then there is a chance that it would hire ten designers from a non-protected classification. If the population of applicants includes members of a protected group, then this hiring method, where candidates are ranked according to capability, could potentially produce a disparate impact.[26] It is easy to understand why law provides an exception to the fashion house for legitimate business needs. The satisfaction of these needs supports its competitive advantage.

---

[21] *See* Office of the Comptroller of the Currency et al., *Interagency Fair Lending Examination Procedures*, app. at 27 (Aug. 2009) (noting burden is on plaintiff in context of lending decisions); *see also* Texas Dept. of Housing and Community Affairs v. Inclusive Communities Project, Inc., 576 U.S. 519 (2015); 78 Fed. Reg. 11459, 11482 (Feb. 15, 2023) (same for mortgage decisions). Placing the burden on divorcees is consistent with other areas of law, including employment decisions. *See* 42 U.S.C. § 2000e-2(k)(1)(A)(ii) (placing burden on plaintiffs to show less discriminatory alternatives to employer's method of decision-making).

[22] *See* Emily Black et al., *Less Discriminatory Algorithms*, 113 GEORGETOWN L. J. 53, 86–87 (2024).

As mentioned earlier, needs must be socially legitimate, but the achievement of business needs can be difficult to measure, which provides employers with substantial latitude to hire as they wish. For instance, companies can hire on the basis of "best fit," a notoriously difficult concept to measure.[27] In lending law, legitimate business needs are also broadly understood. Creditors may generate adverse impacts in lending if their facially neutral efforts are aimed at minimizing credit risk. It is often remarked that disparate impact rules have little bite because, among other things, demonstrating statistical disparity is challenging,[28] but the rules can be substantially weakened in as much as socially acceptable goals are broadly defined. Broader goals present a greater opportunity for disparities to arise. There is an assumption built into the law that the benefits of lawful goal achievement outweigh the costs of indirect discrimination.[29]

---

[23] Random allocation is effective to the extent that every person in the population makes an identical contribution toward achieving a goal.

[24] Of course, clever resource allocators who know that randomization will disfavor a particular group can use random selection in order to discriminate. The point, albeit abstract, is that randomization eliminates intentional discrimination when outcomes are behind a veil.

[25] This is true so long as the population is heterogeneous. Thus, less diverse populations experience less discrimination all other things equal.

[26] Taken literally, a disparate impact means that a protected group receives more benefits than a non-protected group. However, groups differ in number, and law attempts to correct for their size by assuring comparable distributions of benefits. Thus, the Uniform Guidelines of Employee Selection Procedures defines a disparate impact (sometimes used interchangeably with the term adverse impact) as a "substantially different rate of selection in hiring, promotion, or other employment decision which works to the disadvantage of members of a race, sex, or ethnic group." Uniform Guidelines on Employee Selection Procedures, 43 Fed. Reg. 38,290 (Aug. 25, 1978). A rate is said to be substantially different when a non-protected group receives a benefit more frequently than a protected one, where "more frequently" is a magnitude that reaches some threshold set by law. For example, if 80 percent of married applicants and 20 percent of single applicants are hired, then the ratio of single hires to married hires is 20:80, or 1/4. This ratio represents a "substantially different rate." In the past, any rate less than 4/5 represented potential discrimination and could trigger a need to show a legitimate business need. Today, it is widely understood that the 4/5 rule is arbitrary even if occasionally useful. Other statistical measures are typically used (and criticized). Watson v. Fort Worth Bank and Trust, 487 U.S. 977, 994 (1988) (noting that plaintiff must show "statistical evidence of a kind and degree sufficient to show that the practice in question has caused the exclusion of applicants for jobs or promotions because of their membership in a protected group").

In general, anti-discrimination law separates the social welfare effects of discriminatory action into three components: the private goals of the college, lender, and employer; the social costs of the decision-making tool, i.e., discrimination; and the private costs of developing alternatives. As seen in *SFFA*, law determines whether private goals are socially acceptable. If goals are unlawful, then direct and indirect discrimination are prohibited.[30] Otherwise, the rules seek to enable private goal achievement at the lowest private and social cost. The following Sections describe the anti-discrimination rules specific to admissions, lending, and employment in detail, but first, consider further the nature of the goals of a university, lender, or employer.

### *A. Admissions Decisions*

As mentioned above, the *SFFA* court declared race-based admissions programs violate law if they "lack sufficiently focused and measurable objectives," "unavoidably employ race in a negative manner, involving racial stereotyping," and "lack meaningful end points." The holding is expected to undo affirmative action in college admissions. It is easy to see why. Apart from measurable goals, *SFFA* requires that colleges and universities refrain from "unavoidably employ[ing] race in a negative manner, involv[ing] racial stereotyping."[31] The majority noted that using race in a positive manner for one group necessarily (and indirectly) causes a negative effect for another group. Seats in an incoming class are finite, and allocation on the basis of race necessarily creates racial winners and losers.[32] Take, for instance, two students who are otherwise identical except for their race. If, for example, the seat is won for being Native American, then the same seat is lost, albeit indirectly, for not being Native American. If the non-admitted applicant is, say, East Asian, then she loses solely on the basis of her race. It may be true that the college does not directly penalize the student for being East Asian, but this makes no difference if admissions are zero-sum. The penalty is an indirect effect of using race in a positive manner. Even if a school were able to articulate measurable goals, the combination of racial plus factors and a finite admissions policy leads, inevitably, to indirect penalties for members of another race. One solution is to eliminate finite admissions and enroll all applicants.[33] Race could then lawfully be considered, since all students would be admitted and none would be penalized indirectly. However, its consideration would no longer be meaningful. Another solution is to confront the plus factors head-on and simply declare that race can be used as part of an effort, for example, to remedy past wrongs.[34] This approach, of course, was rejected.

[27] *See* ORLY LOBEL, THE EQUALITY MACHINE 48 (2022) (noting the phenomenon).

[28] On the difficulty associated with establishing statistical evidence of disparities, *see Watson*, 487 U.S. at 996 (noting that courts and defendants can challenge a plaintiff's statistical evidence and that plaintiffs must isolate an unlawful method). For a critical history of disparate impact law, including the challenge of establishing evidence of disparity, *see* Ryan H. Nelson, *Substantive Pay Equality: Tips, Commissions, and How to Remedy the Pay Disparities They Inflict*, 40 YALE L. & POL'Y REV.149, 182-190 (2021).

[29] This is easily seen in settings in which the state is the allocator of scarce resources and receives immunity from disparate impact claims. *See, e.g.*, DiCocco v. Garland, 52 F.4th 588 (4th Cir. 2022) (noting that the government defendant initially pled that the ADEA does not provide for a disparate impact cause of action, but later changed its position as the litigation proceeded).

[30] By direct, I mean discrimination in resource allocation that results from the evaluation of prohibited classes and classifications (disparate treatment). By indirect, I mean the discrimination that results from allocating resources on the basis of variables that can substitute for prohibited ones.

Inasmuch as the pursuit of diversity with race variables is unlawful, a university's reasons or intentions for considering race are irrelevant. Scrutiny of means is of no consequence when a goal is prohibited. Suppose instead that a school wishes to pursue a lawful goal, such as providing an education to underprivileged applicants. If the school considers race in its selection decisions, then it is selecting applicants partly on the basis of race, which is unlawful. It is true that an intention to engage in race-based selection can be inferred because the school considers race, but its intentions are irrelevant in the context of an unlawful goal. Consideration of race is simply barred. This

[31] *SFFA*, 600 U.S. at 188.

[32] *Compare id.* at 271-72, *with id.* at 331 (Sotomayor, J. dissenting). The dissent contended that schools can avoid using race in a negative manner when it is just one factor for consideration. From a statistical point of view, this is a difficult argument to make when one considers its indirect effect in a scenario where incoming seats are finite.

[33] *See* Brandon Busteed, *Elite Universities Should Enroll a Million Students*, FORBES (Feb. 21, 2021), https://www.forbes.com/sites/brandonbusteed/2021/02/20/elite-universities-should-enroll-a-million-students/?sh=59c6ca8d4ba7.

[34] *See SFFA*, 600 U.S. at 384 (Jackson, J. dissenting).

scenario is familiar in law. In tort law, or example, strict liability rules ignore mental states.[35]

When a variable is facially neutral, and its assessment is helpful for achieving a lawful goal, but it simultaneously serves as a proxy for race, then it may seem that an assessment of intention can clarify culpability. After all, the variable is doing two things, and one is lawful but the other is not. From a social welfare perspective, however, lawmakers ignore intent and simply minimize the social costs of lawful goal achievement. Whether a decision-tool unintentionally engages in proxy discrimination does not matter. What matters is whether a different decision-tool discriminates less so.[36] Consider the *Tactical Bomber* case.[37] An army that bombs a munitions factory and foreseeably kills 100 civilians may be less morally culpable than a *Terror Bomber* that intentionally bombs a residential quarter and kills the same number of civilians. A social-welfare maximizing lawmaker, however, searches for another method altogether. If the *Tactical Bomber* can use precision-guided missiles and achieve the same goal while foreseeably killing only 50 civilians, clearly the less harmful method is superior. Any moral argument about good versus bad intentions is irrelevant.[38]

Because *SFFA* considers whether a college can admit more or less applicants on the basis of race, it is a "disparate treatment" case.[39] However, schools are additionally prohibited from producing unjustifiable disparate impacts or discriminatory consequences on the basis of race-neutral variables.[40] For instance, if a facially neutral admissions policy scores Asian students lower than others, then the school must justify its scoring method. Universities cannot help but consider applicant features that are correlated with race. Correlations of varying strength can be found everywhere within a population. It is well known that race is strongly correlated with zip code, parental income, and health.[41] But today's statistical models can uncover dozens and perhaps hundreds more, and unmasking techniques are only growing in sophistication. These correlated variables can be used as substitutes, or proxies, for a race variable intentionally or unintentionally.

---

[35] For further discussion, see Deborah Hellman, *Diversity by Facially Neutral Means*, 110 VA. L. REV. 1901, 1929-31 (2024).

[36] *See infra* note 116.

[37] It is discussed in Hellman, *supra* 35, at 38 n. 147.

[38] There is an important exception discussed *infra* at Section III.A, in which a proxy perfectly or nearly perfectly substitutes for a prohibited variable. Knowing and intentional use of a near perfect substitute is identical to engaging in disparate treatment. However, as before, intentions do not matter. Usage of the substitutable variable triggers strict liability.

[39] *See* Department of Justice, TITLE VI LEGAL MANUAL, SECTION VI- PROVING DISCRIMINATION- INTENTIONAL DISCRIMINATION, https://www.justice.gov/crt/fcs/T6Manual6 (accessed Jan. 9, 2023) ("A Title VI discriminatory intent claim alleges that a recipient intentionally treated persons differently or otherwise knowingly caused them harm because of their race, color, or national origin.").

[40] *See* Department of Justice, TITLE VI LEGAL MANUAL, SECTION VII- PROVING DISCRIMINATION- DISPARATE IMPACT, https://www.justice.gov/crt/fcs/T6Manual7 (accessed Jan. 9, 2023). Note, however, that claims must be brought by the state. There is no private right of action for disparate impact claims under Title VI. *See* Alexander v. Sandoval, 532 U.S. 275 (2001).

Regardless of the goal sought, the use of proxies can often lead to disparate impacts. Admissions law will consequently face growing questions about proxies, especially when the evidence of their use is opaque and circumstantial. It seems likely that law will set boundaries on their permissible use. As will be discussed shortly in Part III, one possibility is that if the university can use two algorithms that achieve the same goal with the same amount of success, but one deploys fewer proxies, then that algorithm should be used. Returning to the example above, if a university can justify its scoring method that adversely impacts Asians, then Asian applicants should be able to suggest a better and less discriminatory method, presumably one that does not consider applicant features that correlate with being Asian.

### *B. Lending Decisions*

Lending law prohibits disparate treatment and impacts in decisions over interest rates and the provision of credit.[42] Disparate treatment occurs when lenders treat applicants who possess identical credit-risk factors differently on the basis of membership within a protected class. As in employment decisions, disparate impacts occur when "[a] creditor employs facially neutral policies or practices that have an adverse effect or impact on a member of a protected class."[43] Creditors are absolved if a policy "meets a legitimate business need that cannot reasonably be achieved by means that are less disparate in their impact."[44]

---

[41] *See* Alberto Alesina, Matteo F. Ferroni & Stefanie Stantcheva, *Perceptions of Racial Gaps, Their Causes, and Ways to Reduce Them*, NBER Working Paper No. 29245 (Sept. 2021), https://www.nber.org/system/files/working_papers/w29245/w29245.pdf.

[42] *See* 12 C.F.R. Pt. 1002- Equal Credit Opportunity Act (Regulation B) (amended Aug. 29, 2023) (prohibiting discrimination "in any aspect of a credit transaction"). The Equal Credit Opportunity Act of 1974 (ECOA) prohibits discrimination on the basis of the usual protected classifications of race, color, religion, national origin, sex, marital status, and age, as well as the applicant's use of public assistance programs for income, or a previous good faith exercise of rights under the Consumer Credit Protection Act. *Id.* For claims related to the provision of home loans or home improvement loans, a claim may be pursuant to the Fair Housing Act in addition. *See* The Fair Housing Act, 42 U.S.C. § 3601 *et seq*. The protected classifications under the Fair Housing Act are race, color, religion, sex, national origin, family status, or disability.

[43] 12 C.F.R. Pt. 1002 Supp. I Sec. 1002.6(a)-2.

[44] *Id.*

Unlike admissions, lawful goals in lending are clear. Repayment of a loan is clearly a legitimate business need of the lender.[45] Consequently, creditors can (and do) take into account proxy variables for race and other protected classes when making lending decisions.[46] Variables such as income, credit score, and debt are correlated with race,[47] and lenders routinely assess these variables in order to predict an ability to repay the loan.

Intuition suggests that these proxies are permitted because they strongly correlate with repayment. But what if a new model of repayment behavior is found that exhibits a weaker connection to protected classifications?[48] Law clearly requires the use of the new model in favor of the older, more discriminatory model so long as their private costs are comparable. A lender, for instance, may wish to use a credit model built from its own data to avoid paying for costly credit reports of its customers. But suppose the lender's model heavily weights zip code, a strong proxy for race, and that the model based upon credit score is less discriminatory. The lender might plead that the use of its cheaper, proprietary model is itself a legitimate business need since the model lowers lending costs.[49] The outcome will likely depend on whether business necessity is broadly defined.[50] If the costs of using either model are comparable, then there is a clearer case for using the credit score model because of its lower proxy power.

### C. *Employment Decisions*

Rules against discrimination in lending closely track rules against employment discrimination under Title VII. The Civil Rights Act prohibits both disparate treatment, which occurs when an employer takes action directly on the basis of race or other protected attributes, as well as disparate impacts, which occur when decisions guided by facially neutral parameters lead, nonetheless, to uneven outcomes.[51] Employment discrimination law possesses a number of peculiarities that suggest the underlying justification for a claim might best be developed statistically. First, discriminatory action must be identifiable as a company- or department-wide "policy or practice" as opposed to a number of loosely connected ad-hoc or one-off decisions.[52] Policies and practices must essentially be decision tools that are routinely and broadly applied, which can be understood as demanding a sufficient number of observations for drawing a credible inference of a non-random connection between a policy and a disparate impact.[53] Today it remains unclear whether the use of algorithmic decision-making tools amounts to a policy or practice. On the one hand, delegated discretion given to lower-level company managers is not a policy or practice,[54] and companies can delegate decisions to machines. On the other hand, to the extent that the company defines and controls the algorithm, its decisions are more the result of a routine policy and practice.[55] In any case, this Article proceeds under the assumption that the use of an algorithm suffices as a company policy.

---

[45] *See supra* note 18.

[46] *See* Aaron Klein, *Credit Denial in the Age of AI*, BROOKINGS, (Apr. 11, 2019), https://www.brookings.edu/articles/credit-denial-in-the-age-of-ai/.

[47] *See* SHIFT PROCESSING, *What Is a FICO Score and How Is It Determined?*, (last visited Jan. 18, 2024), https://shiftprocessing.com/credit-score/#race (noting that average credit scores for Asians, Whites, Other, Hispanics, and Blacks is 745, 734, 732, 701, and 677, respectively).

[48] An example is provided *infra* at notes 84-87 and accompanying text.

[49] Another example of increased costs involves an ongoing debate over how much data a financial institution should be required to collect. Consumer Financial Protection Bureau Rule 1071 requires lenders to collect demographic data on potential clients. The stated purpose is to allow government to monitor predatory and abusive lending practices, including discrimination. *Id.* However, the additional data collection (and associated compliance and paperwork) disadvantages small banks and credit unions. *See* S.J. Res. 32, 118th Cong. (2023).

[50] A broad recognition of business needs will include turning a profit at the lowest possible cost in order to maximize the likelihood that the firm remains solvent over time. For additional discussion, see *infra* notes 59-60 and accompanying text.

Second, the discriminatory action must lead to a measurable adverse impact. Establishing such an impact typically involves comparing the effects of a policy across two or more different groups. A disparate impact is said to occur if one group receives a benefit or a detriment at a "substantially higher rate" than the other.[56] "Substantially higher rate" suggests the need to credibly identify a pattern well beyond an ordinary random occurrence. Of course discerning between patterned and random events is best accomplished with elementary statistics.[57] Again, the analysis presented here is agnostic

---

[51] *See* Griggs v. Duke Power Co., 401 U.S. 424, 427, 431 (1971) (declaring that Title VII "proscribes not only overt discrimination but also practices that are fair in form, but discriminatory in operation"). This rule has been codified at 42 U.S.C § 2000e-2(k)(1)(A)(i) (2018).

[52] *See* Wal-Mart Stores, Inc. v. Dukes, 564 U.S. 338, 338 (2011).

[53] A sufficient number of observations is necessary for making a plausible claim of wrongdoing.

[54] *See Wal-Mart Stores*, 564 U.S. at 355-357.

[55] The EEOC has even suggested that the company could be held responsible for algorithms designed and administered by another company. *See* U.S. Equal Employment Opportunity Commission, *supra* note 17.

[56] *See supra* note 26 (discussing the meaning of a "substantially different rate" and the weakening of the four-fifths rule).

[57] Consequently, various statistical tests are used to measure adverse impacts. For example, a Chi-square or Phi-coefficient test is typically used when the employee population is large and Fisher's exact test is preferred when the population is small. Other tests, like a standard t-test, are used to measure the before and after effects of a planned policy. Finally,

about cut-off points, such as four-fifths rules, and proceeds under the assumption that an adverse impact has been shown under existing rules.

Once a plaintiff successfully pleads the presence of a policy and an adverse impact, then the employer may offer a defense that the policy is "job-related for the position in question and consistent with business necessity."[58] As might be expected, there is a longstanding debate over the scope of this defense.[59] On the one hand, any policy that increases profits and helps the firm avoid long-term deterioration and bankruptcy can be understood as a necessity. On the other, if a policy is only modest in cost, then a competitive firm may be able to withstand any loss and continue to subsist or even thrive.[60] Once the scope of necessity is set, if the employer is able to establish that its policy is job-related and consistent with its lawful business necessity,[61] then, as is well known, the plaintiff may offer an alternative policy that satisfies the employer's interests and produces a less disparate

---

logistic regression can be used to measure impacts while controlling for a number of other variables such as geography, job type, performance, education, and seniority.

[58] *See* 42 U.S.C. § 2000e-2(k)(1)(A)(i).

[59] *See* Ian Ayres, *Market Power and Inequality: A Competitive Conduct Standard for Assessing When Disparate Impacts Are Unjustified*, 95 CALIF. L. REV. 669, 685 (2007).

[60] Notice that the first possibility of "any policy that increases profit" possesses the virtues of coherence and easy commensurability, while the ability to subsist over time while incurring losses is speculative and harder to measure. Some have suggested that a third standard of "reasonable risk-adjusted returns on capital" should be used. *See* Ayres, *supra* note 59 at 672. Returns could be derived from beta regressions using the Capital Asset Pricing Model. *Id.* at 672 n.12. However, if competitors outperform the computed returns, say, because of comparatively lower capital costs due to market psychology unmoored from standard valuation metrics, then the less competitive firm may eventually face bankruptcy.

[61] Defendants were required to rebut employers' proffered justifications. *See* Wards Cove Packing Co. v. Antonio, 490 U.S. 642, 659-60 (1989). The onus was later placed on employers with legislation. *See* 42 U.S.C. § 2000e-2(k)(1)(A)(i).

impact.[62] If the employer refuses to adopt it, the plaintiff prevails.[63] Thus, as before in the admissions and lending contexts, if an employer can use two algorithms that achieve the same goal, but one deploys fewer proxies for race or other discriminatory variables, then that algorithm should be favored on the basis of existing rules. Employment discrimination law places the burden on plaintiffs to offer a better algorithm. Ideal burden placement will be discussed at length in Part IV. It is now time to turn to the identification of proxy variables.

## II. PROXY VARIABLES

### *A. Perfect Substitutes for Protected Class Variables*

Begin with the most extreme case in which a facially neutral variable perfectly substitutes for membership in a protected class. In statistics, this is sometimes referred to as a perfectly linear relationship.[64] The presence of one variable predicts the presence of another in all cases. If a facially neutral variable is a perfect substitute for protected class membership, then using the neutral variable is no different, statistically, from using a non-neutral variable like race or marital status.

In life, it is rare to observe variables that serve as perfect substitutes for each other. Different scales of measure, such as Fahrenheit and Celsius, are interchangeable because they represent the same information in different formats, but variables that convey different information routinely differ even if they co-relate. After all, when variables convey identical information, it makes less sense to separate them as representatives of a single phenomenon. Even if rare, perfect substitutes do occur and near-perfect substitutes more so. Consider that perfect substitutes can be asymmetric, so that the presence of variable X predicts the presence of variable Y, but not the opposite. Asymmetric substitutes are more common and frequently appear in medicine and biology. For example, there are genetic and other biomarkers that perfectly predict the presence of a disease. While genetic substitutes matter little for employment discrimination, since employers may not discriminate on the basis of genetic information absent a legitimate need,[65] and the use of biomarkers in college admissions and lending is unheard of (at present), it is easy to imagine the strategic use of other asymmetric substitutes that serve as social markers.

---

[62] *See* 42 U.S.C. § 2000e-2(k)(1)(A)(ii); *see also* Albemarle Paper Co. v. Moody, 422 U.S. 405, 425 (1975) (noting that plaintiffs must demonstrate an alternative policy or practice that serves the employers legitimate purpose, but "without a similar undesirable racial effect."). The cost of implementing an alternative is relevant. Watson v. Fort Worth Bank & Trust Co., 487 U.S. 977, 998 (1988); Furnco Construction Corp. v. Waters, 438 U.S. 567, 577-78 (1978). Plaintiff's bear the burden of production and persuasion with respect to identifying a better policy that still serves the employer's interest. *See* 42 U.S.C. § 2000e-2(k)(1)(A)(ii) (plaintiffs must "make[] the demonstration described in subparagraph (C) with respect to an alternative employment practice, and the respondent refuses to adopt such alternative employment practice"). Demonstration under the statute means "meets the burdens of production and persuasion." 42 U.S.C. § 2000e(m). Finally, subparagraph (C) provides "the demonstration referred to by subparagraph (A)(ii) shall be in accordance with the law as it existed on June 4, 1989 with respect to the concept of 'alternative employment practice.'" 42 U.S.C. § 2000e-2(k)(1)(C). Since *Wards Cove* was decided a day after, the statute effectively overrules its placement of the burden on plaintiffs to rebut the employer's justification for the policy to be a business necessity. The plaintiff nonetheless continues to bear the burden of demonstrating an alternative policy or procedure that is less discriminatory.

[63] 42 U.S.C. § 2000e-2(k)(1)(A)(ii).

[64] SEE WOOLRIDGE, *supra* note 8. In the context of regression analysis, the phenomenon is often referred to as perfect collinearity.

Consider the presence of an accent. Setting aside regional differences, accents can potentially substitute, statistically, for national origin if all people who speak accented English are born outside of the United States. The relationship is asymmetric since many people who are born abroad do not speak English and thus have no English accents. Even so, a dataset that exhibits perfect substitutability of accent for national origin is probably rare. Many people that are born overseas have American parents or school teachers, and they learn to speak unaccented English at a young age. Near-perfect substitutability of accent for national origin is more likely.[66] Suppose that the accent variable substitutes for national origin ninety-five percent of the time and that a university, perhaps aware of this relationship between the two variables, uses accent to routinely deny admission. A disparate impact claim can clearly be brought, but a case can be made for discriminatory intent

---

[65] *See* 42 USC § 2000ff-1.

[66] There are several notable examples of near-perfect substitutes in employment discrimination law. In Rogers v. American Airlines, 527 F. Supp. 229 (S.D.N.Y. 1981), the plaintiff asserted that her cornrow hairstyle was strongly associated with African American culture and a workplace prohibition on that style amounted to Title VII discrimination. The court sided with the defendant. Several states responded with legislation. *See, e.g.*, S.B. 188, 2019-2020 Legis. Sess. (Cal. 2019), https://leginfo.legislature.ca.gov/faces/billTextClient.xhtml?bill_id=201920200SB188. Second, there are a number of disputes over "speak English only" workplace rules, in which the EEOC has said are tantamount to national origin discrimination. [cite]. Finally, in Pers. Adm'r of Mass. v. Feeney, 442 U.S. 256, 279 (1979), despite the fact that 98 to 99 percent of veterans in the relevant population were male a policy of favoring veterans was upheld as non-discriminatory in its purpose.

In addition, the Supreme Court has characterized the location of a home as a substitute for face in its line of racial gerrymandering cases. *See* Shaw v. Reno, 509 U.S. 630, 657-58 (1993); Miller v. Johnson 515 U.S. 900, 905 (1995); Bush v. Vera, 517 U.S. 952, 957-58 (1996); Cooper v. Harris, 581 U.S. 285 (2017).

just as well.[67] Part of its strength lies in the strong relationship between accent and national origin. There is a hidden assumption that a stronger co-relationship is more easily observed, and that its conspicuousness implies a greater likelihood that the university is using it with knowledge or intention.[68] This assumption only increases in power once the university is made aware of the connection, yet continues to use the variable.

### *B. Imperfect Substitutes*

Co-relationships between variables can be revealed with statistics. A standard measure is correlation. In general, correlation measures the presence and strength of a relationship, but says nothing of its magnitude. For example, correlation between the strength of an accent and national origin can be measured by calculating a correlation "coefficient" across a set of data.[69] The calculation returns a number, the so-called coefficient, between 0 and 1.[70] This number says little about the substitutability of accent for national origin because it only measures how tightly the co-relationship is bound, not its magnitude. For example, suppose that the correlation coefficient for accent strength and national origin measures 0.36, suggesting a moderate to strong relationship. This does not imply that a strong accent increases the likelihood that a person is of a particular national origin by 36 percent. In order to test for the magnitude of the accent effect, a different method, such as regression analysis, must be used.

---

[67] As noted in the Department of Justice's Title VI Legal Manual, "discriminatory intent need not be the only motive, a violation occurs when the evidence shows that the entity adopted a policy at issue "'because of,' not merely 'in spite of,' its adverse effects upon an identifiable group." (quoting *Feeney*, 442 U.S. at 279). Department of Justice, *supra* note 39.

[68] *See Bray*, 506 U. S. at 270.

[69] Note that "presence of an accent" can only take two values: yes and no. Similarly, national origin can only take a finite number of values, e.g. all of the political subdivisions of the world, though it can be coded as "American" and "non-American." When a variable takes on a finite number of values, including two like yes/no, it is known as a "categorical" variable as it reveals the presence and absence of categorical states. Categorical variables are important for anti-discrimination law because they are used to represent protected classes. Even age, typically represented with a continuous number, can be coded in groups, e.g. over and under 40. When measuring correlation between categorical variables, a data analyst must be careful to select the appropriate test for measurement. While a Pearson's test may be appropriate for measuring correlation between continuous variables, a Chi-Square test should be used for categorical variables. *See* WOOLRIDGE, *supra* note 8, at 6. Note that other measures such as ANOVA or logistic regression must be used to measure correlation between a categorical and continuous variable.

[70] Some correlation coefficients, such as the popular Pearson's variant, return a number between -1 and 1. This measure is appropriate for variables that take on continuous values like height or weight. As noted *supra* note 69, Pearson's test is inappropriate for categorical variables like accent type or national origin.

Suppose a data analyst builds a regression model for national origin that includes one predictor variable, accent strength, and that diagnostic testing of this model reveals an R-squared[71] of 22 percent. If the model can be trusted,[72] then 22 percent of the variation in national origin is explained by accent strength. R-squared, in this example, measures the strength of the correlation. Clearly, accent strength cannot perfectly substitute for national origin. The R-squared in the regression model would need to measure 100 percent. However, legal tests for discrimination do not require the use of protected class variables nor their perfect substitutes. As long as a policy or practice produces an adverse effect, then its use is potentially prohibited. It seems sensible that proxies should be prohibited if they are easy to identify[73] and their elimination produces a less discriminatory effect while still satisfying the lawful goals of a college, employer, or financial institution. Under these conditions, their elimination would produce an alternative decision-making method that is both less discriminatory and equally effective.

A model that contains many variables will require testing for correlation between each variable and a protected class. In the past, this would have been a time-consuming task, perhaps amounting to an unreasonably costly pursuit of a less discriminatory alternative. Today, large datasets that contain many variables can be tested with relative ease.[74] Proxy variables can be identified and their strength can be measured.

## III. LEGAL TESTS FOR PROXY POWER

---

[71] Depending on whether the variables are continuous or categorical, different regression methods and measures of R-squared must be used. For example, if a dependent variable takes a discrete yes/no value, an analyst will deploy logistic regression. The appropriate measure of R-squared is "McFadden's R-squared," a type of R-squared measure suitable for logistic regression. Like standard R-squared, pseudo measures of R-squared measure how much variation an independent variable explains. For example, if SAT score is regressed on admission decision (a yes/no variable), an analyst would deploy logistic regression. And if the McFadden's R-squared is 0.22, then SAT score explains 22 percent of the variation in admissions decisions.

[72] Trust essentially means that the plots of the residuals exhibit no patterns. If the residuals exhibit patterns, then the result is biased, and the model, and its measure of R-squared, cannot be trusted.

[73] As mentioned *supra* note 62, the cost of implementing an alternative is relevant. *See Watson*, 487 U.S. at 998; *see also Furnco Construction Corp.*, 438 U.S. at 579-80. Thus, the cost of identifying proxies and testing the effectiveness of alternative decision-making tools without them is relevant. Today, these costs are falling. *See infra* note 75.

[74] For an example in which a data analyst checks for correlation across 15 pairs of categorical variables with a single piece of Python code, *see* Ritesh Jain, *Correlation Between Categorical Variables*, MEDIUM (May 31, 2020), https://medium.com/@ritesh.110587/correlation-between-categorical-variables-63f6bd9bf2f7.

### *A. The No Proxy Rule*

The most straightforward and blunt regulation of proxies is to simply prohibit their use. In most cases, prohibition makes little sense given that social welfare is maximized by balancing legitimate goals against the social costs of indirect discrimination and the private costs of developing less discriminatory alternatives. Nonetheless, as is presently discussed, there is a limited case to be made for prohibiting proxies under certain conditions involving perfect or near perfect substitutes. Consider again the context of admissions.

It seems likely, if not certain, that *SFFA* prohibits the use of a race variable in college admissions algorithms. Imagine an algorithm returns a single suggestion to accept or reject the applicant. If the algorithm includes a race variable, then it is suggesting admission partly on the basis of race, which today is barred. The same can be said for any decision-making method. Human admissions committees may not consider race on its own, though a committee (and also an algorithm) may consider how race has brought about other attributes in a candidate, such as the example of courage discussed earlier.[75] From a statistical point of view, and perhaps the Court's, the committee would simply be considering courage. While the analysis to this point has focused on proxies for race and other protected-class attributes, note that here, race serves as a proxy for courage, which is clearly a variable that may be lawfully considered, assuming a lack of discriminatory motive. Thus, the permissible inclusion of race in an admissions decision tool seems to turn on the inclusion's purpose. Is race included to screen candidates on the basis of race, or is it included as a surrogate measure for courage or some other desirable attribute? The Court seems to indicate that inclusion of a surrogate measure would be unlawful because surrogates involve racial stereotyping.[76] Using race as a proxy for courage essentially reveals that the algorithm user believes that members of a particular race possess more or less courage on average. If this belief were not present, then use of the proxy would make no logical sense. Assuming correlation involves racial stereotyping.

From a model building standpoint, the trouble with using race as a proxy for courage is that courage is directly observable. Proxies are generally used (purposefully) when a variable is hard to measure or when direct observation is impossible or too costly. Upon review of a personal essay, the committee or algorithm learns, easily and straightforwardly, of an applicant's courage. Perhaps its connection to race makes the presence of courage more credible. If so, then race is simply a proxy for credible courage, which is also directly observable upon review of an essay. For both of these reasons—stereotyping and the availability of direct observation—it is hard, and likely impossible,[77] for colleges after *SFFA* to lawfully include race in a decision tool regardless of its stated purpose.

[75] *See supra* note 4 and accompanying text.

[76] Stereotyping is impermissible. *SFFA*, 600 U.S. at 211.

Proxies for race, on the other hand, are necessarily used in admissions decisions inasmuch as race is associated with facially neutral variables that are relevant for achieving a lawful goal.[78] If the goal were academic achievement, for example, then SAT score, high school GPA, high school attended, number of AP classes, awards, and other similar variables could be included, even though they correlate with race. But suppose high school is a perfect or near substitute for race for 20 percent of applicants. This is not so difficult to imagine, as many high schools are highly segregated.[79] Suppose, again, that an algorithm returns a single proposal to accept or reject an applicant. This suggests that the algorithm is making decisions partly on the basis of race, albeit indirectly, for 20 percent of the applicants.

Of course this is not evidence of intentional discrimination, but again, if the college were made aware that high school substitutes for race 20 percent of the time, intentional and knowing discrimination appears more plausible. This is especially the case if the high school variable explains a large part of the decision to admit those applicants who attend the segregated schools, say, because the algorithm gives less weight to SAT score, high school GPA, and its other variables.[80] If this were the case, then admissions decisions would be strongly influenced by race by means of its proxy variable: high school. In this example, by contrast, the high school variable likely explains little variation in admissions decisions since other variables which measure the college's goal of maximizing academic achievement are included in the algorithm. Nonetheless, law prohibits the intentional use of plus and minus factors.[81] If the high school variable were to explain any part of the admissions decision, then its use by an admissions committee, which is aware of its perfect substitution for race, could potentially amount to a plus or minus factor, even though perfect substitution is occasional and occurs only 20 percent of the time. If it explains no part of the admissions decision, then there is no reason to use it.

---

[77] Perhaps a situation could arise in which the desirable attribute is not directly observable and measuring it indirectly with race does not involve stereotyping if, for instance, its connection to race is random. But if that were the case, then it would no longer serve as a proxy. There must be some amount of correlation in order for race to statistically represent courage in a model.

[78] As stressed above, this Article takes the lawfulness of goals as given. Whether educational diversity is a lawful goal has been discussed elsewhere. *See supra* note 20.

[79] U.S. GOV'T ACCOUNTABILITY OFF., *Student Population Has Significantly Diversified, But Many Schools Remain Divided Along Racial, Ethnic, and Economic Lines*, at 11 (June 2022), https://www.gao.gov/assets/gao-22-104737.pdf (noting that 14 percent of K-12 students attend "almost-exclusively same-race/ethnicity schools").

[80] As discussed above, one way to measure how much high school explains variability in admissions is to use linear regression analysis and observe the R-squared (or pseudo R-squared).

[81] *See SFFA*, 600 U.S. at 230-231.

Note that the high school variable is less problematic for decisions regarding the applicants who attend the non-segregated schools. Because their schools are not perfect substitutes for race, use of the high school variable in their admissions decisions does not imply knowing use of plus or minus factors. This example raises the possibility of using two algorithms for different groups of people on the basis of high school, which probably constitutes disparate treatment since high school is a (near) perfect substitute for race. Consider nonetheless an edge case in which an applicant pool consists of 40,000 people, and only 100 have attended segregated high schools. The college might believe that the high school variable could help it make better admissions decisions for applicants that attended non-segregated high schools—the vast majority—and that using two algorithms imposes an unreasonable cost. It wishes to use a single algorithm that contains the high school variable, even though that variable substitutes for race in 0.25 percent of its applicant pool, a small amount. Nonetheless, if high school partly explains admissions decisions and the college is aware of its perfect substitution for race, then use of the variable impacts the entire applicant pool insofar as admissions are zero-sum. Its use would likely be prohibited.

Perhaps case law will evolve to strongly disfavor the high school variable since its deployment can implicitly signal a knowing use of race. The signal comes through because of its near perfect association with race 20 percent of the time, but also because other variables with a weaker relationship to race, such as SAT score and high school GPA, can accomplish the same goal as the high school variable purports to do. The present example may carry the features of a disparate impact claim, but the suggestion is that a claim for intentional and knowing disparate treatment can be made on the basis of using a perfect or near perfect substitute when less perfect substitutes are available.[82] The argument proceeds by process of elimination. If other available variables can accomplish the goal of the university, employer, or bank, and the only impact of a proxy is to screen for race (or another protected class), then there is little justification in law to use it. Knowing use of a superfluous proxy amounts to discrimination.

### B. *The Comparative Minimum Proxy Power Standard*

[82] Near perfect substitutes can be treated as perfect ones inasmuch as nearness gives rise to an inference of knowing discrimination.

Efficient prohibition is likely rare. In most cases, perfect substitutes such as the example of the high school variable do not exist, and universities will narrowly tailor their algorithms and other decision tools in order to use variables that partially substitute for race. Narrow tailoring can be accomplished by minimizing the use of partial substitutes. Similarly, the use of partial substitutes in lending and employment contexts may be lawful inasmuch as a business necessity can be shown and less discriminatory alternatives are unavailable. In these instances, the lawful use of proxies will turn heavily on the availability of better alternatives. Note, too, that the use of perfect substitutes for protected class variables like sex and age may be permitted for business needs. Their lawful use will turn on the availability of alternatives just as well.

Over time, it should be expected that growing computational power will unveil previously hidden predictors of behavior and bring about new alternatives for making decisions. The arrival of new alternatives will generate less discriminatory bases for decision-making at a lower or comparable cost. For example, in a recent study it was shown that digital footprint variables were more predictive of loan default than credit score for small consumer loans.[83] Researchers examined 250,000 credit transactions completed on Wayfair.de, a shopping website similar to Amazon.[84] As expected, a credit bureau score did a good job predicting repayment and default, but digital footprint variables were more sensitive, especially as credit score declined.

The digital footprint model contained three variables related to the applicant's computer, three related to their email address, one that accounted for how the user found the Wayfair website (i.e. through an advertisement or by entering the web address into a browser), and one that captured the time of day credit is requested.[85] While the variables are facially neutral, they

---

[83] *See* Tobias Berg et al., *On the Rise of FinTechs—Credit Scoring Using Digital Footprints*, FDIC CTR. FOR FIN. RSCH. Working Paper 2018-04 (Sept. 2018).

[84] Small loans of up to € 400 were extended to 254,808 customers so that they could complete their online purchases. *Id.* at 39. If the amount exceeded € 100, Wayfair requested a credit bureau report. *Id.* at 9. The amount had to be repaid within 14 days. If repayment was late, the customers received up to three notices before entering default and transfer of the claim to a collection agency. *Id.* at 9.

[85] Values for computer variables were desktop, tablet, mobile, or not tracked; values for operating system were Windows, iOS, Android, Macintosh, other, not tracked); and values for email host were Gmx – partly paid, Web – partly paid, T-Online – affluent customers, Gmail – free, Yahoo – free older service, Hotmail – free older service, and other. The researchers additionally deployed combination variables such as Android User who applies for credit late in the evening. *See id.* at 33.

The default rate for desktop users was 0.74 percent; tablet users 0.91 percent; and mobile users, an astonishing 2.14 percent. For users whose device was not tracked by Wayfair, the default rate was 0.88 percent. Android users defaulted at the highest rate (1.79 percent). Mac probably are correlated to race and other classes protected by lending law by means of mediating variables.[86] Consider their relationship to income. Suppose iPhone users, on average, earn higher incomes than Android users. Because income is associated with race, it is likely that accessing Wayfair with a comparatively cheaper Android device, on average, is associated with race. Similarly, using an odd email name such as crimson_tide_18 instead of one's first and last name can be associated with unemployment since company email addresses usually contain names. Because unemployment is additionally associated with race, odd email addresses will correlate with race in this example. Other associations between the digital footprint variables and variables correlated with race can be imagined.

Under the assumptions that using a particular model is a "policy or practice" and managing the risk of loan repayment is a legitimate business need, Wayfair's less discriminatory alternative consists of assessing an applicant's ability to repay with a workable model that deploys a collection of variables that indirectly discriminates the least when compared with other models. Suppose that, for a given dataset, the digital footprint model produces a loan repayment accuracy rate of 95 percent. For the sake of illustration, suppose further that a credit score model consisting of three variables—credit score, income, and age—predicts loan repayment with identical accuracy. However, the credit score model's proxy power for race is 0.08. In other words, race, albeit indirectly, explains 8 percent of the variation in loan repayment. In contrast, suppose the digital footprint model's proxy power is 0.05. The digital footprint model in this example is the least discriminatory alternative.

Intuitively, computation of proxy power considers the correlation strength of a variable as well as the variable's importance for explaining variation in an outcome like loan repayment or academic performance.[87] By

---

users defaulted at the lowest (0.69 percent). Users of paid email hosts default at the lowest rates. Users of free and older services default at the highest. Inclusion of name, no inclusion of numbers, and usage of lower case were all associated with lower default rates. The opposites were associated with higher default rates. Later times were associated with higher default rates. Finally, an email error was associated with high default. All default rates can be found *id.* at 40.

[86] The Wayfair study was carried out in Germany, but suppose for the moment that a digital footprint model can predict loan default in the United States.

[87] As mentioned earlier, comparing the R-squared of two linear regressions, one with say credit score, and the other with age, offers a straightforward method for assessing their relative importance. In the case of multiple regression analysis, where many variables are regressed on loan default, the analyst can add one variable at a time and assess the difference in R-squared with and without the additional variable. There are a number of pitfalls that can lead to inaccurate inferences, which are beyond the scope of this Article. For an overview, see JIM FROST, REGRESSION ANALYSIS: AN INTUITIVE GUIDE FOR USING AND INTERPRETING LINEAR MODELS (2020).

combining correlation strength and explanatory importance, one can measure how much a prohibited variable indirectly explains an outcome.[88] Intuitively, these values can be computed for each variable in a model and added together, for instance, to measure the total proxy power of a model. The model with the lower number will produce comparatively less indirect discriminatory effects.[89] This approach, while intuitive and illustrative of the proxy power concept, can produce errors in measurement. In some instances, the indirect effect of race on loan repayment can be accurately assessed with a simple "Cumulative R-squared" analysis. Take the digital footprint model. Regress the model's variables on repayment and observe its R-squared. Say R-squared is 65 percent. Now add race. If R-squared increases to 70 percent, then race explains 5 percent of repayment in the digital footprint model. If one carries out the same analysis with the credit score model, and finds that

---

[88] Taking the product of correlation strength and explanatory power represents an intuitive, but potentially inaccurate, measure of the indirect impact of race on repayment. Another possibility is to estimate a model using all variables, including the protected class attributes. Once estimated, the coefficients for the protected class attributes can be set to zero, so that only the remaining variables (and their coefficients) are used for computing the predicted outcome for each individual in the dataset. Under certain conditions, this method eliminates indirect effects. *See* Devin G. Pope & Justin R. Syndor, *Implementing Anti-discrimination Policies in Statistical Profiling Models*, 3 AM. ECON. J. 206, 206 (2011); *See also* Anya E.R. Prince & Daniel Schwarcz, *Proxy Discrimination in the Age of Artificial Intelligence and Big Data*, 105 IOWA L. REV. 1257, 1313 (2020) (providing further discussion). Direct effects can then be subtracted from total effects in order to measure indirect effect of a model, which can then be compared.

Other debiasing models build on Pope & Syndor. *See, e.g.*, Crystal S. Yang & Will Dobbie, *Equal Protection Under Algorithms: A New Statistical and Legal Framework*, 119 MICH. L. REV. 291, 346-48 (2020) (developing a model that identifies and partitions correlated and uncorrelated effects of a proxy variable and then removes the correlated effects only). Yang and Dobbie develop a second model that infers an uncorrelated effect by benchmarking the predictive effect of a variable across a group that is assumed to be treated fairly. *Id.* at 348-50. Estimated coefficients of the fairly treated group are then applied to the unfairly treated group. *Id. Compare* Talia B. Gillis, *Orthoganalizing Inputs*, CSLAW '24: Proceedings of the Symposium on Computer Science and Law (March 2024) (noting that accurate identification and removal of uncorrelated effects is difficult because pervasive correlations reduce the stability of identification and hence accurate removal of correlated effects). In general, Tallis suggests that algorithmic colorblindness is a myth in practice. *See also* Kristen M. Altenburger & Daniel E. Ho, *When Algorithms Import Private Bias into Public Enforcement: The Promise and Limitations of Statistical Debiasing Solutions*, 175 J. INST'L & THEORETICAL ECON. 98, 98 (2019), which shows that the validity of the Pope & Syndor approach often requires assuming that the protected class variables are independent of other predictors. Note that the approach taken here completely side-steps the need to debias a model. Models that consist of facially neutral variables are simply ranked based upon lowest proxy power.

[89] It will produce comparatively less indirect discriminatory effects for the race variable. When a model implicates multiple protected classes, computation for each will need to be completed. *See infra* notes 98-99 and accompanying text.

race explains more than 5 percent of repayment, then the digital footprint model is the less discriminatory alternative.[90]

Consider a final example of an employer. The employer receives thousands of resumes each day for its open positions. It wishes to develop an algorithm to screen the best candidates. Over time, the employer has learned how much employees contribute to profit—similar to how today's sports teams measure the impact that individual players have on a team's winning percentage. Because a potential hire must work well with the employer's existing workforce, the algorithm favors the status quo by accounting for a set of features that captures the habits and capabilities of current employees. Thus, the algorithm consists of two broad categories of variables: those related to the individual candidate and those related to combinations of candidate and workforce features. The combination variables can be understood as a measure of how well the candidate will fit into the existing workforce culture. Individual attributes and cultural fit both contribute to profit. The model captures nothing else.

Suppose the screening algorithm scans incoming resumes and assigns values to the individual and cultural variables. Individual variables include things like education and work history, military service, and hobbies. Cultural variables consist of how well the candidate's individual variables fit with the workforce's. These are more complicated and account for things like how many existing employees share the candidate's hobbies or school history and other measures of the likelihood of rivalry and cohesion. On the basis of these variables, each candidate is assigned a score in terms of dollars. These dollars represent the expected amount of profit that a candidate will generate to the employer each year. For example, if the candidate graduates from the University of Michigan, spent several years working in operations management, served in the Army Corp of Engineers, and lists piano as a hobby, then the increase to expected annual profit is $200,000. If the workforce strongly prefers working with a colleague who has no operations experience, then the same candidate will score $180,000.

---

[90] Statisticians will recognize this measure as the "semi-partial r-squared" or "part r-squared" of the race variable. The difference in r-squared between the two models with and without race reveals how much variation in the goal is uniquely or incrementally accounted for by race. The inference is that a model which explains less rather than more incremental variation in goal achievement (when adding race) exhibits less proxy power. One must be careful with this approach because R-squared may be inaccurate, especially if the model is biased. *See supra* note 72; FROST, *supra* note 87. In addition, models that contain many variables tend to yield higher values of R-squared. If one is comparing a model with few variables to a model with many, then adding race may cause a substantial increase to R-squared when added to the model with few variables as opposed to the model with many. Whether or not Cumulative R-squared is a good measure depends on context and requires careful analysis.

Now suppose the employer collects data on undergraduate major and that this variable is highly correlated with sex.[91] When undergraduate major is removed from the model, however, the same candidate scores $200,000.[92] The model, without this variable, appears to be a less discriminatory alternative for screening potential employees. Note that in this example, removal of the undergraduate major variable matters little. The candidate will still receive the job offer if $200,000 is higher than her competitors' predicted profit increases. But if she scores $150,000, then the employer might successfully claim a business necessity defense. Either the proxy does not impact the employer's decision, or if it does, the proxy is necessary for achieving a goal. It may seem, therefore, that the minimum proxy power rule accomplishes little. The absence of an effect is less likely, however, as the number of candidates grows. Suppose the employer must hire 100 people. With and without the undergraduate major variable, the model projects an average profit increase per candidate of $200,000. The new hires increase total projected profit by $20,000,000. When the undergraduate major variable is included, the algorithm suggests hiring 60 males and 40 females. Without the proxy variable, the algorithm suggests a 50-50 balance. In either case, total projected profit of $20,000,000 remains identical,[93] but the employer can no longer so easily claim business necessity. The minimum proxy power rule requires the removal of the offending variable.

In general, an analyst can compare the mean accuracy of a model that predicts changes to profit with and without the sex variable.[94] If mean accuracy is identical for the two models, then removal of name reduces the proxy power of sex.[95] There are at least two potential inaccuracies to this approach. First, it necessarily compares a potential candidate's profit contribution to the aggregate contributions of past and current employees. If the candidate possesses some unique attribute that is associated with sex that contributes to profit, then this attribute will be lost on the model and the employer may assert that assessment of sex is a business necessity. The loss in accuracy is probably smaller for larger workforces since there is a higher chance that the candidate will not possess a unique and meaningful attribute that is unaccounted for in the model. Second, and potentially more serious, is that the model's accuracy relies on how well future employees reflect past and current employees. The employer can only build a model at a specific moment in time. At that time, the employer's testing data necessarily consists of past and current employees.[96] The inferences one can draw regarding the ability of future employees to contribute to profit are necessarily tied to a predictive model built from employee behavior and capabilities measured earlier in time. To the extent that the future resembles the past, any inaccuracy due to change will be low. In sum, as long as future employees look like past and current employees, comparing the mean accuracy of profit prediction on a representative testing dataset can be a viable method for comparing models. If mean accuracy is identical with and without the name variable, then the employer can achieve its business necessity with less proxy power for sex. Furthermore, the removal of *any* variable correlated with sex, however weakly, that also does not reduce the mean accuracy of a model can safely be removed. The model's overall accuracy will not suffer, and the removal of variables correlated to sex will deliver a comparatively lower amount of indirect discrimination.[97] This is true for proxies of any protected-class attribute.

[91] *See* CATHERINE HILL, CHRISTIANNE CORBETT & ANDRESSE ST. ROSE, WHY SO FEW?: WOMEN IN SCIENCE, TECHNOLOGY, ENGINEERING, AND MATHEMATICS *passim* (2010) (noting correlations between sex and undergraduate major).

[92] Put statistically, if sex were regressed on additional profit, R-squared would measure zero.

[93] This occurs because the model's other attributes are used to explain more of the variation in projected profit, or perhaps because the employer is encouraged to add another variable like courage (as in the example in the Introduction), which generates a better fit of the variables to projected profit. Strategies for inducing employers to experiment with different models are discussed *infra* in Part IV.

[94] Specifically, the model predicts profit for existing employees contained within a testing dataset, where actual profit has already been observed. The model, for instance, may predict $1,950,000 for an employee who increased profit by $2,000,000. Mean accuracy of a model can be computed by calculating the average differences between the actual and predicted values. If mean accuracy is identical with and without a variable correlated to sex, then the model without that variable represents a less discriminatory alternative. *See* GARETH JAMES, DANIELA WITTEN, TREVOR HASTIE & ROBERT TIBSHIRANI, AN INTRODUCTION TO STATISTICAL LEARNING: WITH APPLICATIONS IN R 330 (1st ed. 2013).

[95] Again, this is true so long as sex explains some variation in outcomes. If it does not, then the proxy power of the variable is zero even though it is highly correlated with sex.

In practice, indirect discrimination on the basis of multiple protected classifications may be implicated by the use of a decision tool. For example, in addition to the association of undergraduate degree to sex, the model might include zip code, which is associated with race. A model that minimizes sex may increase proxy power for race. One way to address this complication is to select the model that yields the average minimum proxy power in comparison to others. In the example above, the algorithm suggests hiring 60 males and 40 females when undergraduate major is included. Suppose inclusion of military history causes the algorithm to suggest a balanced 50-50 pool of males and females, but leads to fewer job offers for Native Americans. A less discriminatory alternative may attempt to minimize average proxy power by assessing the expected number of candidates that possess a protected-class attribute. It is easy to see how this could work with an edge case. If no Native Americans apply, then the military history variable can be safely included and generate a sex-balanced pool. As more Native Americans apply, however, a difficult trade-off must be made. The same challenge will arise if the algorithm presents trade-offs over multiple protected classifications, for example, if a candidate pool consists of people that originate from many different nations.

[96] The employer could use synthetic data, but this approach is not completely data-driven. It requires making assumptions about unobserved counterfactuals that could produce yet other inaccuracies, which may lead the employer to assert the need to account for sex. Note, too, that once an employee is hired, all of the cultural variables will require updating if the new employee brings even slightly different preferences for future employees, since the new employee's preferences will likely impact and reconfigure those of the existing workforce whose preferences are dependent on the newcomer by construction.

[97] As stressed in the example and footnote 95, if the addition of a proxy variable generates no increase to the explanatory power of the model, measured through, say, its contribution of additional R-squared, then the proxy variable has no indirect discriminatory effect on employee selection. However, if the additional explanatory power is zero, then there is no statistical reason to include the proxy variable. It provides no additional predictive power to the model. Employers will remove it on their own if only to save on data collection and processing costs.

Finally, imagine the case where removal of a proxy variable reduces mean accuracy of a model only slightly. Suppose the use of the name variable leads to an average increase in annual profit of $100 per employee but leads to dramatic sex imbalances in the workforce. Although the employer can put forward a business necessity rationale for using the proxy, intuition seems to point toward a narrowing of its legitimate goal, at least a small amount, if its method for achievement generates elevated proxy power. However, making the tradeoff necessarily recognizes that a social goal is not legitimate. Any deviation, however small, can be cast as an entirely new goal. For example, if the previous goal were profit-maximization, then the new goal might be profit-maximization less one percent of last year's sales.[98] For small differences in goal achievement, perhaps a standard of "practical equivalence" will evolve as a means to side-step difficult determinations of lawful and unlawful goals.[99]

### C. *The Capped Proxy Power Rule*

---

[98] As highlighted above, this article takes lawful and unlawful goals as given, and sets aside questions about their optimal scope. Of course the argument for removing proxies could be strengthened by narrowing legitimate profit maximization. On narrowing this goal, see Ayres, *supra* note 59, at 672.

[99] One may be tempted, from a social welfare perspective, to permit the tradeoff absent any equivalence. A different approach, put forward in the following Section, is to recognize that small deviations around goals can be the result of differences in measuring the impact of variables. Those differences are to be expected given the degrees of freedom available to analysts when computing the effects of a model. As discussed below, caps on proxy power economize on the administrative and enforcement costs associated with the precise computation of proxy power, which can justify the elimination of an offensive proxy that only slightly increases profit.

As rules prohibiting or limiting the use of proxies begin to develop, litigants and lawmakers will observe imperfections in measurement. While computing correlation is usually straightforward and possible, it is difficult to accurately measure a variable's explanatory or predictive power. Assumptions must be made about the goodness of a model, how well its training dataset represents the real world, and whether its inferences can continue to justifiably be made in the future. Nonetheless, as illustrated above, some models are better than others, even if distinguishing between close substitutes is hard. It should be expected that in close cases, proxies will be permitted since credible identification of a less discriminatory alternative will not be possible. In easy cases, where a model and its application are discernibly reliable, training data is clearly representative of the real world, and the domain of decision-making is relatively stable so that reliable inferences can be made in the future with past data, and it will be easier to justify limitations to proxies.

Some domains, such as university admissions and lending, may be more stable than others (like employment in a particular industry). If so, then it may make sense to use caps on proxy power. For close cases in which models cannot be credibly distinguished with statistical tools, a threshold could be used to provide at least some baseline guidance for model selection. Suppose a university has two models for admitting students so that the models maximize the achievement of a lawful goal. The first deploys 10 variables that correlate with race and the second deploys 5. However, proxy power for selecting students with the first model is only slightly lower than the second. If so, then it is unlikely that the model with 5 proxies represents a less discriminatory alternative with certainty given the difficulties and degrees of freedom associated with the measurement of proxy power.[100] But consider that over time and with experience, universities may, for example, learn that differences in proxy power in terms of cumulative R-squared of 5 percent or less represents a fair margin for error. Suppose when race is added to the five-proxy model, R-squared increases by 4 percent. When race is added to the ten-proxy model, R-squared increases by 6 percent. In this instance, the models can be credibly compared, and the five-proxy model would be favored. But when comparing two models that both fall under the 5 percent threshold, any differences would be ignored.

Five percent might emerge as a reasonable cap given the industry and other features of the model's environment. This approach highlights the well-known tradeoff between rules and standards. As rules, proxy caps reduce the administrative costs of carrying out detailed analyses on a case-by-case basis but could increase errors in identifying true indirect discrimination inasmuch as the 5 percent cap is too high or low. The broader point is that statistical imperatives are more suited to rules-based architectures in stable as opposed to dynamic domains.[101] When environments are stable, efficient caps are more likely to emerge. In practical terms, the state lowers its administrative costs by sidestepping the need to judicially assess evidence, including, most importantly, the computation of proxy power across a number of decision-making alternatives. Rules can lower private costs, too, inasmuch as a college, lender, or employer is encouraged to stop searching for and developing alternative decision-making tools, e.g. permutations of a model with and without specific variables. In the example above, defendants may lawfully stop searching for a better model, for example, once they discover a model whose R-squared only increases by 5 percent when adding race. Over time, as models grow in sophistication or life changes, the threshold could fall to 3 percent or less or increase to 6 or more.[102]

[100] For instance, it would be difficult to reach the threshold of a preponderance of evidence standard.

### *D. Unmeasured Goals*

Of course, proxy power cannot be minimized if a defendant does not state its goals or if its goals are subjectively interpreted. In mathematics, minimization makes little sense without an objective function. The function precisely defines what is being optimized and allows for a comparison among the methods used to achieve optimization. So it is with proxy discrimination. For instance, a better way to achieve a 90% bar passage rate within an admitted class is to use an algorithm with fewer or less powerful undesirable proxy variables. A comparison amongst algorithms is possible because the goal of "90% bar passage" is clearly defined. Compare the *SFFA* context where Harvard sought to "train future leaders, acquire new knowledge based on diverse outlooks, promote a robust marketplace of ideas, and prepare engaged and productive citizens."[103] Measuring the achievement of these goals is challenging, as they raise additional questions and their success is subjective. For example, are certain types of leaders more desirable than others, such as business leaders versus political leaders? Additionally, what defines a leader? Fortune 500 CEOs are surely leaders, but what about a Boy Scout Patrol Leader who significantly influences young scouts? Answers often involve social perceptions of prestige, which itself is variable across time and sub-groups of people. An admissions committee can therefore increase its discretion and decisional latitude when its goals are less clearly defined. But most importantly, without a clearly defined goal and consensus on how it can be measured, comparative minimization of proxy power cannot occur. This is because there is no goal to establish the parameters for evaluating the desirability of various methods for achieving the goal.

---

[101] *See* Frank Fagan & Saul Levmore, *The Impact of Artificial Intelligence on Rules, Standards, and Judicial Discretion*, 93 S. CAL. L. REV. 1, 1 (2019), https://southerncalifornialawreview.com/wp-content/uploads/2020/01/93_1_Levmore.pdf.

[102] As noted *supra* note 99, caps present the additional advantage of reducing proxies in instances in which they help a defendant reach a lawful goal only slightly, but generate large disparities across groups. Because the cap is based upon collective experiences of similarly situated defendants, the threshold will, in many instances, be tethered to something approximating average goal achievement. Compliance with a cap will suppress proxies that only provide a small increase to goal achievement to the extent that small increases would be used to surpass average goals not reflected in the cap. For example, a university might boost academic achievement by a small amount when surpassing the 5 percent threshold, but this strategy, and its associated use of additional proxy power, would be barred.

[103] *SFFA*, 600 U.S. at 214.

Another obstacle to reducing proxy discrimination in real-life practice is that people often do not state their goals. For example, even if an employer articulates a broadly shared objective like profit maximization, it does not necessarily calculate the impact of each hire on profit. As noted above, considerations such as a good fit with the existing workforce can be attributed to a profit motive. However, this reduction is rarely made explicit let alone calculated.

There are at least three answers to these awkward complications. First and most obviously, in many contexts, goals are explicit and sufficiently objective from the outset. If this is the case, then a consensual approach to their measurement can be easily achieved. Lending is a good example. Lenders seek to maximize repayment. The goal is clear, stated up front, and objectively measurable. Goals are also clear when mandated by statute or limited to a range of permissible considerations, which ensures that a specific goal must essentially be pursued. This is often the case with government contracting statutes. A municipality or other locale generally seeks to award contracts to the lowest bidders even if it may consider other factors such as quality of workmanship and reputation.[104] The overall objective to minimize cost, with some leeway for quality, is *ex ante* explicit and easily measured. In these instances, where goals are clear, the proxy power of alternative decision-making tools can be compared.

Second, new law can mandate that goals be commensurable and objective. The *SFFA* court noted that race-based college admissions programs must pursue objectives that are "'sufficiently measurable to permit judicial [review]' under the rubric of strict scrutiny."[105] Goals such as training excellent leaders, promoting a robust and diverse marketplace of ideas, and preparing engaged citizens are standardless, that is, there are no acceptable measures of what constitutes an "excellent leader", a "robust and diverse idea marketplace", or an "engaged citizen." By contrast, a prison's stated objective of reducing inmate violence is measurable because the standard is clear. A decrease in violence within the prison population constitutes a reduction and can easily be measured by counting instances of violence at the prison. Thus, the Supreme Court deemed the temporary racial segregation of inmates permissible because it was aimed at achieving that measurable goal.[106]

[104] *See, e.g.*, Tex. Loc. Gov't Code Ann. §§ 252.042-.043 (West 2023).

[105] *SFFA*, 600 U.S. at 214 (*quoting* Fisher v. University of Tex. at Austin, 579 U.S. 365, 381 (2016). The Court further noted that "'[c]lassifying and assigning'" students based on their race, "'requires more than . . . an amorphous end to justify it.'" *Id.* (quoting Parents Involved in Community Schools v. Seattle School District No. 1, 551 U.S. 701, 735 (2007)).

The Court could readily extend the "sufficiently measurable" requirement beyond race-based admissions programs subject to strict scrutiny to those receiving intermediate or rational-basis review. After all, any level of judicial review applied to an admissions program necessitates analyzing the connection between the program's methods and its intended goal. How can this relationship be adequately examined if a goal remains unclear?[107]

---

[106] Johnson v. California, 543 U. S. 499, 512–513 (2005).

[107] The *SFFA* court highlighted this point by clarifying that "sufficiently measurable" requires articulating a connection between goals and methods. *SFFA*, 600 U.S. at 214-18. For instance, to measure educational diversity, a university must accurately assess the racial composition of its incoming class using appropriate racial categories. If a college lacks a category for West Africans, the connection between educational diversity and the college's race-based admissions program becomes tenuous, exemplifying an under-inclusive method for achieving diversity. Furthermore, methods can be over-inclusive or arbitrary. An over-inclusive program might have a general category for Asians without sub-categories for East or South Asians. Such an overbroad admissions method could result in a class deemed diverse by its designers but composed of, for instance, 100 East Asians and no South Asians. An arbitrary category example is "Hispanic," which typically includes individuals in the United States with ancestors from Spain or Latin America. If an admissions method fails to define "Hispanic," it could result in 100 percent of the admitted Hispanic students having Spanish ancestry. This approach clearly does not facilitate measurable diversity. Imagine a sports league aiming to promote safe play by adjusting its rules to reduce injuries but ignoring certain types of injuries when assessing the effectiveness of those adjustments.

In the past, the Supreme Court permitted universities to essentially plead "trust us, we are using race to enhance educational diversity." The Court relied on admissions committees to sensibly use race to benefit some applicants over others in alignment with the university's goals and pursuit of excellence in the classroom and society. This trust led the Court to recognize a "tradition of giving a degree of deference to a university's academic decisions" as long as that deference was "within constitutionally prescribed limits." *Id.* (quoting Grutter, 539 U.S., at 328); *see also* Miller–El v. Cockrell, 537 U. S. 322, 340 (2003) (noting that "deference does not imply abandonment or abdication of judicial review"). However, as emphasized in *SFFA*, a judge cannot provide meaningful deference to admissions decisions that are not sufficiently measurable and concrete to permit judicial review. *Id.* Returning to the sports analogy, it would be like a judge being asked to defer to the sports league's method

While extending the "sufficiently measurable" standard to intermediate and rational basis review would certainly sharpen goals, plaintiffs would still need to embody a classification. But many proxy variables such as zip code, high school attended, and parental income cannot be understood as classifications. Law might treat proxies as classifications, but this approach is not intuitive, and is inconsistent with the concept of facial neutrality as elaborated in disparate impact jurisprudence. There is a better way to demand *ex ante* explicit and measurable goals as presently discussed.

Consider the employment context. As previously mentioned, employers do not routinely state their goals when hiring or taking adverse action against employees. Consequently, plaintiffs will find it challenging to compare various methods of goal achievement and minimize proxy power in practice. However, plaintiffs can sometimes make reasonably accurate guesses about the employer's goals, such as profit maximization, workplace cultural fit, reducing turnover, fostering workforce competition, and so on. Based on these hypothetical goals, plaintiffs can propose a less discriminatory alternative to employee selection—one with less proxy power.

A judge can determine if the plaintiff's suggested goal or set of goals reasonably explains the employer's objectives and allows the case to survive summary judgment on that basis. If the case proceeds, then the burden can shift to the defendant to explicitly state its goals and their relative importance. At this point, the employer can propose a "better" method of its own, again, one that comparatively exhibits lower proxy discrimination. This approach to clarifying goals can be applied not only in employment contexts, but also in admissions and other settings.

## IV. BURDENS AND SECRECY

### *A. Burdens of Production*

Within the contexts of lending and employment, current law places the burden of finding a less discriminatory alternative on the plaintiff.[108] In the past, prior to the arrival of complex algorithms, this may have appeared sensible. A plaintiff could come forward with an alternative test for strength, aptitude, or personality at comparatively little cost. Today's algorithms can deploy hundreds of variables or more that require numerous data for testing. Creating algorithms can be costly. If lenders and employers can produce less discriminatory alternatives at a lower cost, then it may appear sensible to shift

---

for improving safety when the league had no category for recording concussions. Any level of scrutiny, including rational basis, requires a logical connection between goals and means.

[108] *Wards Cove*, 490 U.S. at 659; 42 U.S.C. § 2000e(k)(1)(A)(i) (2018).

the burden of production to them.[109] After all, measuring mean accuracy with or without a variable, for instance, should be a relatively simple and low-cost step for an employer to take. There are several problems with this approach. Most obviously, it places the decision to stop searching in the hands of the defendant, even when the defendant's incentives for inefficiently terminating a search are high. As already mentioned, a search might be privately costly, especially when there are many variables to consider and large datasets to test. In addition, when the private benefits of discrimination are high, a defendant will be encouraged to be dishonest. Some commentators have suggested that law should impose a duty on the defendant to carry out a search that is reasonable in scope.[110] Although a reasonableness limitation could reduce private search costs, it will simultaneously increase public enforcement costs. Determination of reasonableness will likely require the dissection of a complicated evidentiary record. In many cases, a factfinder will need to closely examine the algorithm and data to determine if the defendant's decision to stop searching for an alternative is reasonable.

In easy cases, where variables are few and data is accessible, both the plaintiff and defendant's search costs will be low. Here, it makes even less sense to place the burden on defendants since plaintiffs can engage in low-cost searches of their own. The key, from an efficiency perspective, is to create an environment in which both parties, in all situations, can easily search for less discriminatory alternatives. Algorithmic competition represents one method. In addition to reducing enforcement costs, this approach can substantially increase the likelihood of discovering a decision-tool that exhibits the lowest possible proxy power while still attaining a legitimate goal.[111]

A competition begins by setting aside testing data. This data must adequately represent the defendant's own data used for decision-making. Representativeness may be accomplished, for example, by building a synthetic testing set that mimics the features of a defendant's typical candidates for admissions, loans, or jobs. The parties then construct their own algorithms, and the one that achieves the plaintiff's goal while exhibiting the least amount of proxy power wins.[112] Note that both parties will need to collect their own training data, and that the defendant may possess a cost advantage if this data has already been collected from admissions, loan, and employment applications associated with its business. However, many high-quality, public datasets for training algorithms are readily available[113] and will continue to grow over time. In addition, the public's access to adequate data can be expanded with new disclosure practices that further the construction of synthetic training sets.[114]

[109] *See* Black et al., *supra* note 22.

[110] *Id.* at 99-101.

[111] *See* Saul Levmore & Frank Fagan, *Competing Algorithms for Law: Sentencing, Admissions, and Employment*, 88 U. CHI. L. REV. 367, 367 (2021).

[112] Ideally, neither plaintiff nor defendant can see the testing data until after they have constructed their models. Hidden, or "lock-box," testing data reduces overfit and the plaintiff's associated effort to create a model favorable to her, but not representative of the relevant population of candidates of the defendant. *See id.* at 380-87.

Inasmuch as the costs of enforcing a duty to carry out a reasonable search are high, and competition costs continue to fall, placing the burden on plaintiffs to offer a less discriminatory alternative will continue to make good sense. Algorithmic competition is a comparatively cheaper search device for the state. It also represents a more effective method for revealing less discriminatory alternatives when defendants cannot be trusted and easily held to account. It is true in some employment contexts that a plaintiff's cost of developing an alternative may be greater than simply searching for another job, but this is likely not true when plaintiffs are treated as a group. The aggregate search costs of employees who choose to search for other jobs will likely be high in comparison to the development of a less discriminatory alternative that can be used for all candidates. Thus, the development of the algorithm in this special case presents a collective action problem that can be solved by providing usual incentives such as class actions and contingency fees.[115] In addition, a philanthropist might develop an algorithm that can be adapted and used at a low cost by plaintiffs. The point is that the private costs of algorithm development can be overcome. So long as competition is a viable option, plaintiffs should continue to bear the burden of production. Equipped with algorithms of their own, they can challenge the defendant's algorithm or any opaque method for allocating scarce resources, including secret human decision-making and black-box algorithms. Attention will now be directed toward this last possibility.

### *B. Opaque Decision-Making*

It is difficult to prove intentional discrimination because of its location. It often occurs in places that are difficult for the plaintiff to observe, for instance, behind closed doors, through an email chain on a private server, or in an employer's mind. Disparate impact theory makes it easier to detect discrimination by allowing plaintiffs to focus on an entire population. While individual discrimination may be difficult to detect, aggregate discrimination can more easily be seen when disparate impacts between sub-groups are revealed, even when intention is hidden and no overt and discernible decisions are made on the basis of protected class membership. These disparities occur by means of unobserved intentional discrimination or the use of proxies.

---

[113] *See Datasets*, KAGGLE, https://www.kaggle.com/datasets.

[114] *See* Michal Gal & Orla Lynskey, *Synthetic Data: Legal Implications of the Data-Generation Revolution*, 109 IOWA L. REV. 1087 (2024).

[115] In employment discrimination context, current demand for arbitration imposes substantial hurdles to class formation. In addition, heightened pleading standards may make it difficult to reach the stage of competition. As data is made available, including by means other than litigation, the more likely plaintiffs can meaningfully compete.

This Article has suggested that the use of proxies can and should be controlled. As illustrated above, proxy identification and measurement of their explanatory power can be accomplished when a plaintiff has access to data. The development of less discriminatory alternatives can be accomplished through competition and model refinement. In most of the examples discussed above, the college, lender, or employer uses an algorithm for decision-making. An algorithmic decision-making tool consists of computer code that can be confronted by plaintiffs. A defendant's use of proxies can, consequently, be observed by examining the code.[116] Suppose instead, decisions are made by a committee of humans, such as an admissions committee or board of directors. Because the decisions are in the hands of humans, the use of proxies is, in all likelihood, less directly observable. It is true that the admissions committee can keep records and notes of its decision-making process and a board meeting will surely produce minutes, but the undocumented use of proxies will take place inasmuch as their deployment is not recorded, perhaps taking place only in a decision-maker's mind either consciously or unconsciously.[117] In this case, reasons for decisions are often difficult to discern, but it remains possible that a transparent algorithm represents a less discriminatory alternative when compared to a group of human decision-makers or even an opaque algorithm. At first glance, it may seem difficult to compare the two because the proxy power of variables used in an opaque setting is unobservable or prohibitively costly to observe.[118] The plaintiff's algorithm may be generating an identical result with low proxy power, but it seems difficult to say that it is generating this result with comparatively lower proxy power. This is not the case.

---

[116] For algorithms that use millions of variables, it can be more costly to identify how variables are being used. The task, of course, is not impossible and is usually carried out with standard hypothesis testing. More opaque algorithms simply require more resources to reveal how decisions are being made. Note that competing algorithms sidestep the need to precisely locate and identify indirect discriminatory inputs. Outputs can simply be compared. Another approach is to compel the revelation of inputs with law. *See, e.g.*, 29 U.S.C. 626(f)(1)(H) (providing that employers must disclose age data when negotiating waivers of age-discrimination claims in reduction-in-force scenarios).

[117] The use of "black-box" algorithms can be imagined along the same lines, though as noted above, interrogation is a matter of cost. *See id.*

[118] Detailed examination of the human decision-makers is possible, perhaps with affidavits, testimonies, and tests for bias, but this type of examination is costly and almost certainly more costly than examining a decision model.

Consider the example of employee promotions. Suppose a human resources committee meets every few months to promote middle managers to senior managers. The goal, as before, is to maximize the profit to the company, and each candidate's promotion is assessed on that basis. Candidates are discussed during a committee meeting. At the end of the meeting, each candidate is promoted on the basis of a majority vote. Suppose in March, 300 candidates receive a vote. Two hundred seventy-five are under the age of 40; 25 are over. All of the candidates over the age of 40 are denied a promotion. Suppose discrimination is found on the basis of disparate impact, but the employer offers profit maximization as a legitimate business necessity. The defendant would like to offer a less discriminatory alternative, but cannot observe the HR committee's decision-making process. Say the defendant offers an algorithm that attempts to represent the inputs to a standard HR committee's promotion decision. If the model produces a candidate pool that predicts the identical level of projected profit with more balance across the two groups of employees under and over the age of 40, then the plaintiff has found a less discriminatory alternative. In this way, plaintiffs can deploy transparent algorithms to compete with opaque decision-tools. Although the projected profit of the employer's decision-tool must be disclosed for comparison, and parties may disagree over how to project profit, these impediments can be resolved—the latter perhaps by resorting to industry or another form of consensus. The key point is that decision-making processes and bases can remain hidden. Testing data, perhaps offered by the defendant or assembled synthetically, remains a prerequisite. But its associated cost is likely lower than the cost of the errors that will be produced by the defendant if asked to stop searching for less discriminatory alternatives on its own. From a social welfare perspective, law should allow plaintiffs to succeed when they come forward with an algorithm that produces identical private benefits to the employer and reduced magnitudes of indirect discrimination, regardless of the quality of the variables and other means used by the employer. In other words, the character of a policy or practice should matter little unless it is justified by a legitimate business necessity.[119]

If focus is exclusively placed on the lawfulness of the method used by the employer (and not the adverse outcome), then the standard of comparative minimum proxy power may, in some cases, require a trivial modification if it is to be successfully used to enhance social welfare. To illustrate, consider a university that deploys a hidden decision-making process, whether by means of a black-box algorithm or opaque human deliberation, and applies this inscrutable process to select its incoming class. Upon observing its admissions selections, the university then retrofits a fictitious decision-making process that exhibits low proxy power in order to defend its choices from scrutiny. Disgruntled plaintiffs can only succeed if they manage to find an algorithm that presents an even lower magnitude of proxy power. But this search will be difficult, if not impossible, because the university has retrofitted an inoffensive process to its selections.

---

[119] One way of thinking about policies and practices is in terms of discriminatory intent. *See* Hellman, *supra* note 35. Note, however, when considering the social costs of proxy discrimination, it does not matter whether a policy produces a discriminatory effect intentionally or unintentionally. *See supra* notes 34-38 and accompanying text. So long as a goal is socially acceptable, lawmakers simply minimize the social costs of its achievement.

There is reason to think that the university can successfully retrofit. After all, it possesses numerous data on applicants and can search for unexpected predictors such as the first letter of an applicant's last name or some atypical combination variable such as volunteer work at a food bank and a high school grade in a particular class. However, the university faces some constraints. Most obviously, the more peculiar and unordinary are its variables, the less connection they will have toward reaching a goal such as academic success. It is surely possible, empirically-speaking, that the university may be unable to find a sufficiently retrofitted algorithm with low proxy power that is still able to achieve its goals. But suppose that it can successfully retrofit. Minimum proxy power is no longer an effective device for separating the good algorithms from the bad. Another tool is needed. Ideally, it should be focused on reducing the ability to retrofit. One method is to limit the number of variables that can be used to make decisions. If a plaintiff's algorithm achieves the identical goal, with identical proxy power, but with fewer variables, then its algorithm should be favored. Again, this raises the question of what to do when a goal is only slightly reduced by deploying lower proxy power or fewer variables, but perhaps as mentioned earlier, learning can occur, and efficient caps to the acceptable number of variables will emerge. Another method that avoids the tradeoff is to require the university to announce its selection algorithm at the beginning of the year, before it makes its admissions decisions, or commit to using the same algorithm over a number of years. A pre-commitment would make it difficult to retrofit.

## CONCLUSION

The subtle nature of proxy variables in decision-making tools presents a significant challenge to anti-discrimination law. These variables, while seemingly neutral, can indirectly lead to discriminatory outcomes, particularly in university admissions, lending, and employment decisions.

This Article has suggested that managing proxy variables can best be accomplished by minimizing total proxy power and ensuring decision-making processes are narrowly tailored. This method is necessarily comparative. Over time and with learning, lawmakers may be able to set effective caps on proxy power to the extent decision-making environments remain static. In less understood and fast-changing environments, ongoing comparison will yield the fewest errors.

A key aspect of these rules is the challenge of determining who should bear the responsibility for developing less discriminatory alternatives. This Article has suggested that plaintiffs can play a crucial role in maintaining checks and balances, provided they have access to necessary testing data. Competition ensures a more transparent and accountable system where decision-makers are encouraged to continuously refine their methods to reduce indirect discrimination. In general, the Article encourages critical examination of the overt and covert biases in decision-making processes as well as the degrees of freedom available to those who can effectively wield secrecy.